\documentclass{jfm}
\usepackage[latin1]{inputenc}
\usepackage{graphicx}
\usepackage{amsmath}
\usepackage{natbib}
\renewcommand\Re{\mathrm{Re}}
\newcommand\e{\mathrm{e}}
\newcommand\be{\begin{equation}}
\newcommand\ee[1]{\label{#1}\end{equation}}
\title{An alternate composite representation\\ of the velocity profile\\ in the zpg turbulent boundary layer}
\author{Paolo Luchini}
\affiliation{Università di Salerno, DIIN, 84084 Fisciano, Italy}
\begin{document}
\maketitle
\begin{abstract}
A composite representation of the turbulent boundary-layer velocity profile is proposed, which combines a recently determined accurate interpolation of the universal law of the wall with a simple analytical expression of the smooth transition of velocity to a constant value in the outer stream. Several examples are given of application of this representation to DNS and experimental data from the literature, and a conjecture is offered for the asymptotic approach of the velocity to its constant inviscid value.
\end{abstract}
\section{Introduction}
Coles' (1956) uniform representation of the mean velocity profile as the sum of a law of the wall and a law of the wake, after appearing on Vol. 1 of the Journal of Fluid Mechanics quickly became a standard in the analysis of the turbulent boundary layer, both for purposes of its theoretical description and for the interpolation of experimental data and characterization of their properties. Coles' velocity profile can be written as
\be
u=\left\{\begin{array}{ll}
f(z)+\Pi W(z/\delta)\quad & \text{for}\quad z\le \delta\\
f(\delta)+2\Pi & \text{for}\quad z\ge \delta
\end{array}\right.
\ee{Coles}
where both the wall-normal coordinate $z$ and streamwise velocity $u$ are expressed in wall units (\textit{i.e.}, they are nondimensionalized using the fluid's viscosity $\nu$ and friction velocity $u_\tau=\sqrt{\tau_w/\rho}$, and so will be understood hereafter even if the traditional $^+$ is omitted),
and the wake function $W(Z)$, with $Z$ denoting the outer coordinate $z/\delta$ and $\delta$ itself expressed in wall units, is normalized so that $W(0)=1$, $W(1)=2$; for it Coles proposed $W(Z)=2 \sin^2(Z\pi/2)$. $f(z)$ may be construed to be the complete law of the wall or just its logarithmic portion depending on the range over which \eqref{Coles} is to be applied.

Coles' law provides a family of composite velocity profiles with two free parameters (or three if the dimensional friction velocity $u_\tau$ is counted in addition to the dimensionless external velocity $U_\text{ext}=f(\delta)+2\Pi$ and thickness $\delta$),  which can be fitted to a set of measured velocity data in order to extract its boundary-layer thickness and/or shear velocity, and has ever since been widely adopted for this purpose. Yet \eqref{Coles} was soon recognized to miss the smooth asymptotic approach of the boundary layer to its external constant velocity (so called ``corner defect''), and Coles himself later recommended (Coles 1968) that the shear velocity and boundary-layer thickness should be determined from a restricted fitting of the velocity profile excluding both a range of $z$ near the wall and a range near $\delta$. The difficulty is intrinsic in the logarithmic behaviour of the overlap velocity, which diverges as the wall-normal coordinate tends either to zero or to infinity, and does not match the finite value of the external velocity unless the boundary layer is truncated at a finite thickness (or the wake function is allowed to diverge in the opposite direction).

More recently \cite{Monkewitz} performed an extensive survey of a large number of data sources for the zero-pressure-gradient (zpg) turbulent boundary layer, confirmed the practical adherence of all these profiles (with exceptions ascribed either to presumable imprecision or to inadequate initial conditions) to a two-parameter family of curves, and proposed an alternate composite representation of the multiplicative rather than additive kind which avoids the artificial truncation at a finite thickness; they achieved this result by combining Padé interpolations of suitable order with Euler's exponential integral.

\section{A compact, uniformly valid composite representation}
Purpose of the present note is to illustrate yet another multiplicative composite representation, which in addition to providing similar advantages with a simpler expression, offers an unexpected hint at the asymptotic behaviour of the wake region as it wanes into irrotational flow. Let us take it for granted that the velocity profile $u(z,\Re)$ (where the Reynolds number $\Re$ may for definiteness be the momentum-thickness Reynolds number $\Re_\theta=\theta U_\text{ext}/\nu$) assumes for $\Re\rightarrow \infty$ at constant $z$ the universal form of a law of the wall, $u\rightarrow f(z)$. In a boundary layer, for $z\rightarrow \infty$ at constant $\Re$ it trivially asymptotes to the constant external velocity $u\rightarrow U_\text{ext}$ (which in the case of \eqref{Coles} was $f(\delta)+2\Pi$). The simplest composite expression we can imagine is then a weighted sum of these two asymptotes:
\be
u = wf(z)+(1-w)U_\text{ext}
\ee{composite}
where $w(Z)$, with $Z=z/\delta$, is some yet unknown (hopefully monotonic and well behaved) weight function that goes from $w=1$ for $Z=0$ to $w=0$ for $Z\rightarrow \infty$ over a characteristic thickness scale $\delta$ (conceptually similar but numerically unrelated to the one in \eqref{Coles}).

For any single given velocity profile, \eqref{composite} can be made exact by simply inverting it to get the appropriate $w(Z)$; thus we can start by looking at what $w(Z)$ looks like in typical cases. For the law of the wall we shall adopt the analytical interpolation derived by \cite{EJMB}, repeated here in a notation that can be cut-and-pasted in most computer programs or scripts:
\begin{multline}
\tt f(z)=-(7.374+(0.4930-0.02450*z)*z)/(1+(0.05736+0.01101*z)*z)*\\
\tt exp(-0.03385*z)+log(z+3.109)/0.392+4.48
\label{wall}
\end{multline}
or, for $z\ge 200$, just $f(z)=\log(z)/0.392+4.48$. We recall from \cite{EJMB} that the deviation of \eqref{wall} from its own logarithmic asymptote is non-monotonic, crosses 0 at $z\simeq 23$ and attains a maximum of $+2.5\%$ at $z\simeq 42$; thus it may actually be sufficient to assume $z\ge 30$ as suggested in many classical textbooks, or even $z\ge 20$, for the validity of the logarithmic law provided one is willing to accept an error of this order of magnitude, and does not attempt to evaluate the slope of $f(z)$ in the region $20\le z\le 200$.
For a comparison of \eqref{wall} with the interpolation provided by \cite{Monkewitz} and other historical expressions, the reader is referred to Figure 31 of \cite{EJMB}.

\begin{figure}
\centering
\includegraphics{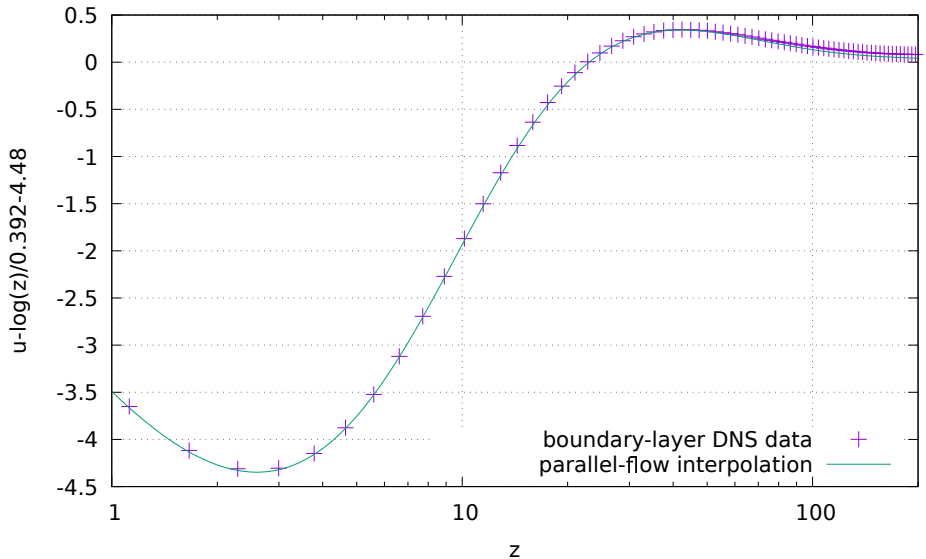}
\caption{Comparison between the actual velocity profile of the boundary-layer DNS of \cite{Jimenez4} at $\Re_\theta=6500$ and the analytical interpolation \eqref{wall} of the law of the wall, obtained by \cite{EJMB} for pressure-corrected parallel flow, both represented as deviations from one and the same logarithmic law to visually amplify any possible difference.}
\label{walldeviation}
\end{figure}
On the other hand, a comparison of \eqref{wall} with a typical boundary-layer profile is provided here in Figure \ref{walldeviation}, which displays the deviation of the mean velocity from the log law (namely the difference $u-log(z)/0.392-4.48$) for the $\Re_\theta=6500$ boundary-layer DNS of \cite{Jimenez4} as compared to \eqref{wall}. This figure ought to rule out any doubts that \eqref{wall}, derived by \cite{EJMB} from an interpolation of parallel-flow data after the pressure-gradient correction of \cite{PRL}, applies just as well to the zpg boundary layer.

\begin{figure}
\centering
\includegraphics{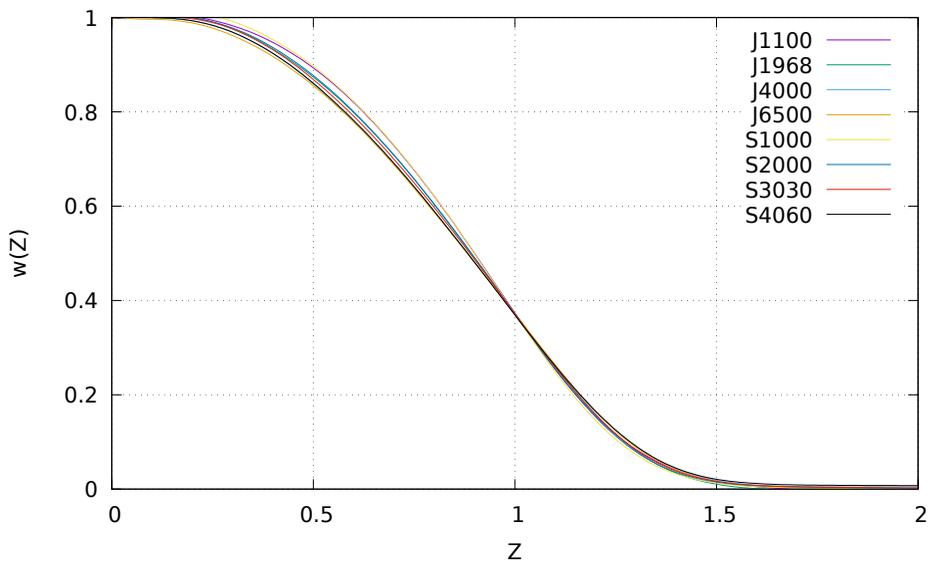}
\caption{Weight function $w(Z)$ as extracted from boundary-layer profiles of different authors at several Reynolds numbers by inverting \eqref{composite}. Curves are marked by a letter 'J' for \cite{Jimenez1,Jimenez2,Jimenez3,Jimenez4}, and 'S' for \cite{Schlatter}, followed by the value of the momentum-thickness Reynolds number $\Re_\theta$.}
\label{extraction}
\end{figure}
Armed with $f(z)$ from \eqref{wall}, we can now go back to the composite expression \eqref{composite}. Figure \ref{extraction} reports several curves for the weight function $w(Z)$ as extracted from the numerical simulation results of \cite{Jimenez1,Jimenez2,Jimenez3,Jimenez4} and \cite{Schlatter}, each normalized on a thickness such that $w(1)=\e^{-1}$. Though not identical, these curves are indeed close enough to each other that they can reasonably be interpolated by a single monotonic, satisfactorily smooth function.

\begin{figure}
\centering
\includegraphics{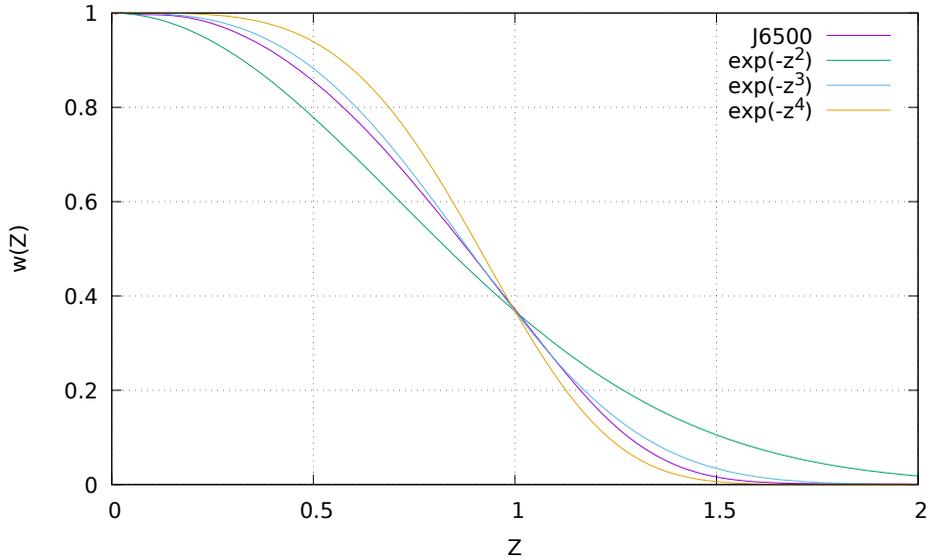}
\caption{$w(Z)$ as extracted from the DNS result of \cite{Jimenez4} at $\Re_\theta=6500$, compared against a gaussian function and against its modifications with different exponents. Behaviour is definitely not gaussian, but a cubic exponential provides a reasonable fit.}
\label{notgaussian}
\end{figure}
In fact the shape of this function looks familiar, and might at first sight be mistaken for a gaussian, but Figure \ref{notgaussian} points out that a gaussian, the exponential of $-Z^2$, is not a good fit. However a little trial and error shows, in the same figure, that the exponential of $-Z^3$ does fit within an error comparable to the dispersion of Figure \ref{extraction}, in a manner that looks more convincing than an accidental coincidence. This is the interpolation we presently propose:
\be
w(Z)=\e^{- Z^3},
\ee{exp3}
which compares well to Coles' $\sin^2$ wake function as to simplicity, but without any corner singularity in $0\le Z < \infty$.
While \eqref{exp3} is very unlikely to hold literally, its good fit does bear two immediate consequences: one is that the combination of \eqref{composite}, \eqref{wall} and \eqref{exp3} provides a very compact practical interpolation of a turbulent boundary-layer velocity profile without corners; the other is a suggestive cue that, although an asymptotic trend deduced from empirical data can never be taken for certain, $\e^{-Z^3}$, and not some other exponential, may be the actual asymptotic law with which the mean velocity profile of the boundary layer approaches its inviscid value. Let us consider each of these consequences in turn.

The folding together of \eqref{composite} and \eqref{exp3} produces
\be
u = f(z)\e^{-(z/\delta)^3}+\left[1-\e^{-(z/\delta)^3}\right]U_\text{ext},
\ee{interp}
\begin{figure}
\centering
\includegraphics{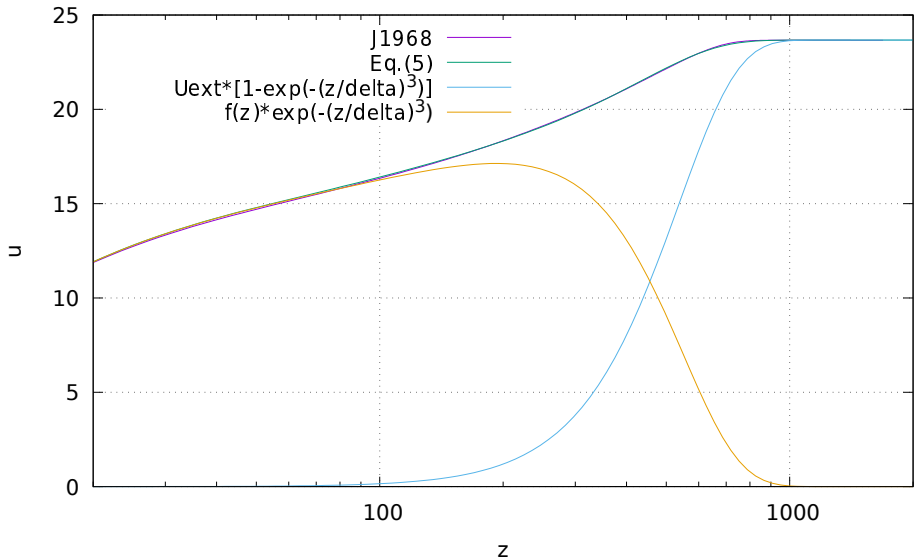}
\caption{Example fit of \eqref{interp} to the DNS data of \cite{Jimenez1,Jimenez2,Jimenez3} at $\Re_\theta=1968$. Also shown are the separate addends of \eqref{interp}, highlighting where the wall and inviscid behaviour is respectively achieved.}
\label{fitexample}
\end{figure}
an expression with two free parameters, $U_\text{ext}$ and $\delta$, which can be used just like Coles' $\Pi$ and $\delta$ to fit an empirical profile and determine its boundary-layer thickness (and, trivially, external velocity). For instance, a least-square fitting of \eqref{interp} to the $\Re_\theta=1968$ data of \cite{Jimenez2} produces a velocity profile graphically undistinguishable from the original (Figure \ref{fitexample}). We note that the fit has been straightforwardly performed on the whole dataset with no exclusions.

\begin{figure}
\centering
\includegraphics{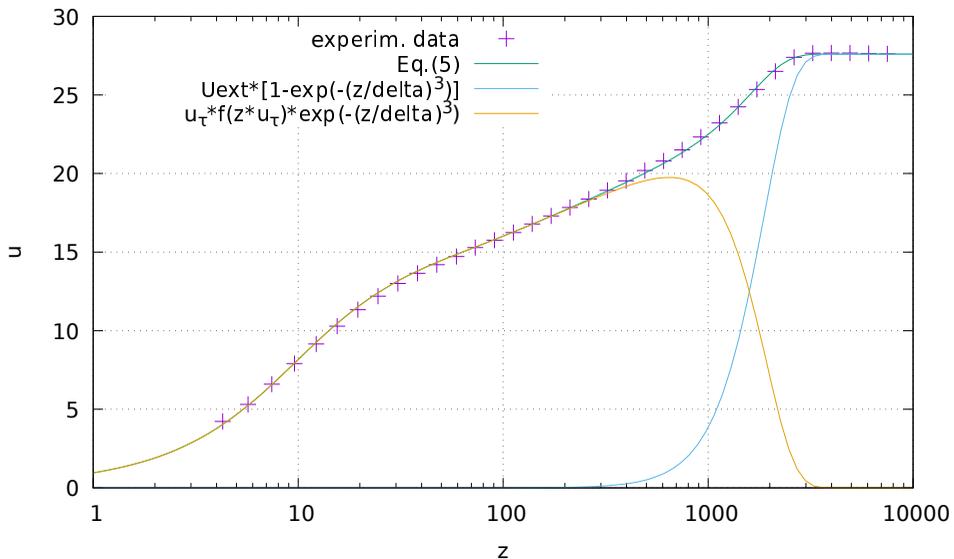}
\caption{Example fit of \eqref{interp} to the experimental data of \cite{Osterlund} at $\Re_\theta=8600$. The shear velocity $u_\tau$ is best-fitted from the data themselves.}
\label{3param}
\end{figure}
\begin{figure}
\centering
\includegraphics{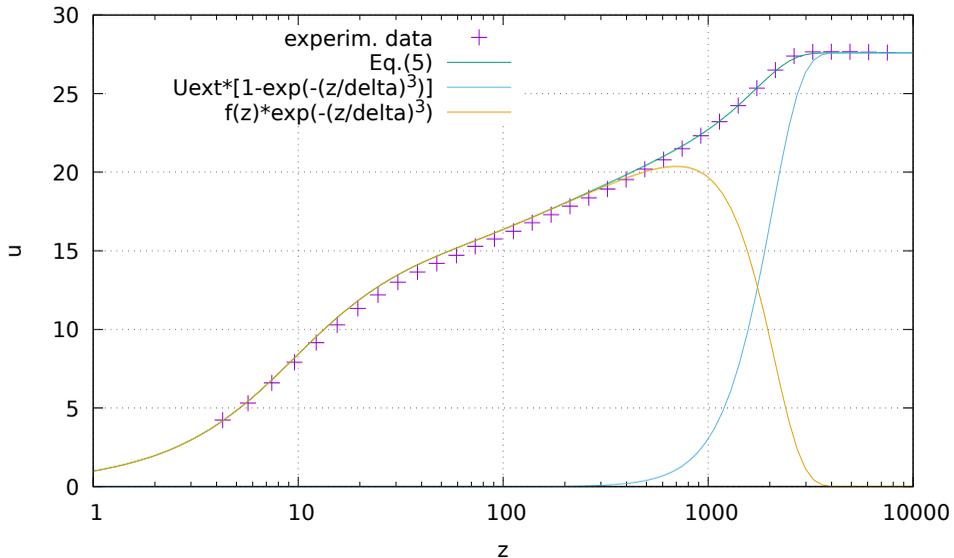}
\caption{Example fit of \eqref{interp} to the experimental data of \cite{Osterlund} at $\Re_\theta=8600$. The shear velocity $u_\tau$ is kept from the original reference.}
\label{3param0}
\end{figure}

Alternately, if $u_\tau$ is also regarded as an unknown parameter, \eqref{interp} can be seen as an expression with three degrees of freedom and used to estimate the wall shear from the velocity profile, in a sophisticated version of Clauser extrapolation. Doing so for the %(randomly chosen) 
experimental profile SW981127K of \cite{Osterlund} at nominal $\Re_\theta=8634.38$ produces the example in Figure \ref{3param}, with a small but possibly significant $-2\%$ correction to $u_\tau$ with respect to the original value (Figure \ref{3param0}). Which one of the two estimates of $u_\tau$ is to be preferred may be open to discussion, or perhaps even be irrelevant, but again we stress that these figures are produced as a fit of the whole dataset with no exclusions.

\section{Outer behaviour and Clauser similarity}
Just like Coles' original formula, the composite formula \eqref{interp} can be separated, if so desired, into its inner and outer behaviour. For $z\ll \delta$, in the inner wall layer, both formulas trivially reduce to $f(z)$ as they are designed to, and hardly any additional remark is needed.

\begin{figure}
\centering
\includegraphics{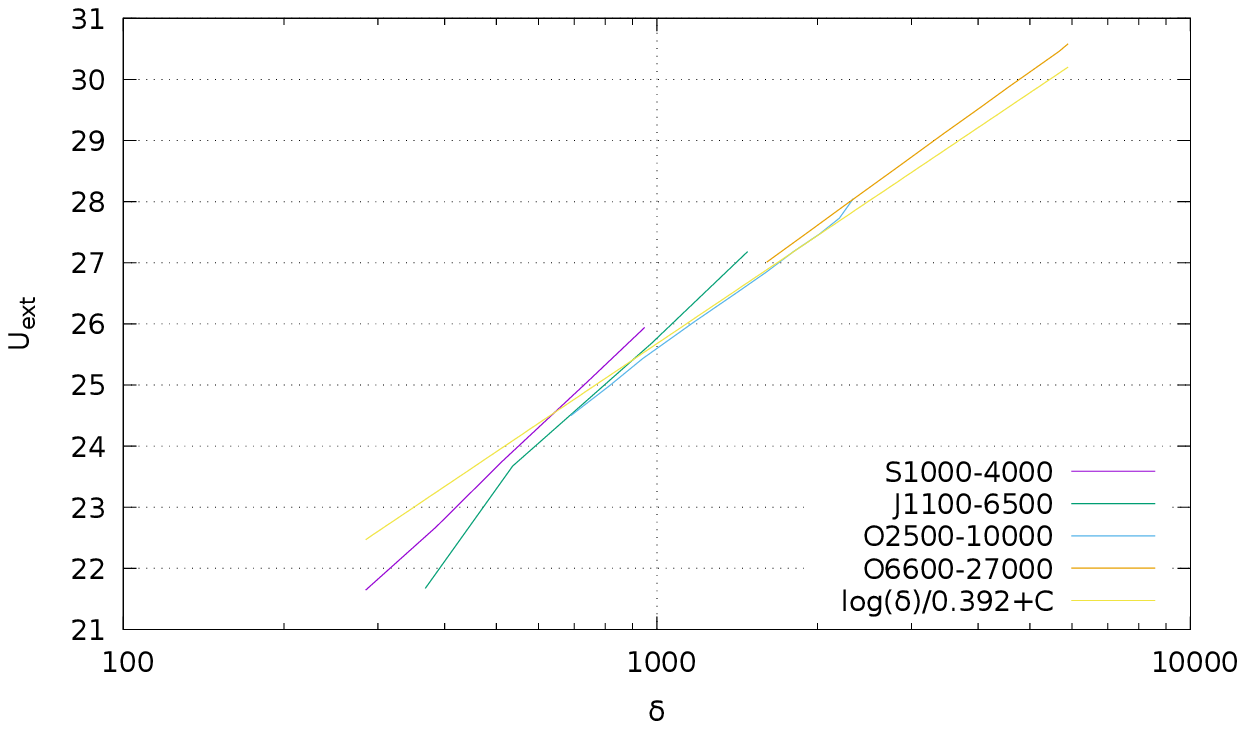}
\caption{$U_\text{ext}$-$\delta$ relationship in the DNS data of 
\cite{Schlatter}, denoted by an initial 'S', \cite{Jimenez1,Jimenez2,Jimenez3,Jimenez4}, denoted by 'J', and the experimental data of \cite{Osterlund}, denoted by 'O'.}
\label{deltaplot}
\end{figure}

For $z\gg 1$ (say, $z\ge 200$), in the outer defect layer, $f(z)$ may be replaced by $\log(z)/\kappa + B = \log(Z)/\kappa + \log(\delta)/\kappa+ B$, where we use $\kappa=0.392$ and $B=4.48$ in agreement with \eqref{wall} and \cite{PRL}, and thus \eqref{interp} becomes
\be
u = U_\text{ext}+\left[\log(Z)/\kappa + \log(\delta)/\kappa+ B-U_\text{ext}\right] \e^{-Z^3}.
\ee{outer}
Clauser's similarity (or equilibrium) regime, generally expected to occur as the Reynolds number (and with it the thickness $\delta$ expressed in wall units) becomes bigger and bigger, requires $U_\text{ext}-u$ to be a function of $Z$ only. This requirement is achieved here if in this limit
$U_\text{ext}\simeq \log(\delta)/\kappa + C$, constant $C$ being in fact the same that would be identified as $2\Pi+B$ in \eqref{Coles}. Whether Clauser similarity is achieved in any given set of numerical or experimental data can easily be ascertained in the present context by plotting $U_\text{ext}$ as a function of $\log(\delta)$, after extracting both parameters from a fit of \eqref{interp} to the data. An example such plot is given in Figure \ref{deltaplot}.
As may be seen, whereas Clauser similarity is conceivably well obeyed in the experiments, the $U_\text{ext}$-$\delta$ line has a significantly different slope in both sets of numerical simulations. Whether this is an effect of Reynolds number (but notice that the slope is different even where the Reynolds ranges overlap) or a forewarning of some other discrepancy is open to further investigation.

\section{Conclusion: a conjecture on asymptotic behaviour}
The success of \eqref{interp} as a uniform interpolating formula leads to the conjecture that $\e^{-(z/\delta)^3}$ may in fact be the appropriate asymptotic behaviour of the velocity defect $U_\text{ext}-u$ at the outer edge of the zpg boundary layer.
\begin{figure}
\centering
\includegraphics{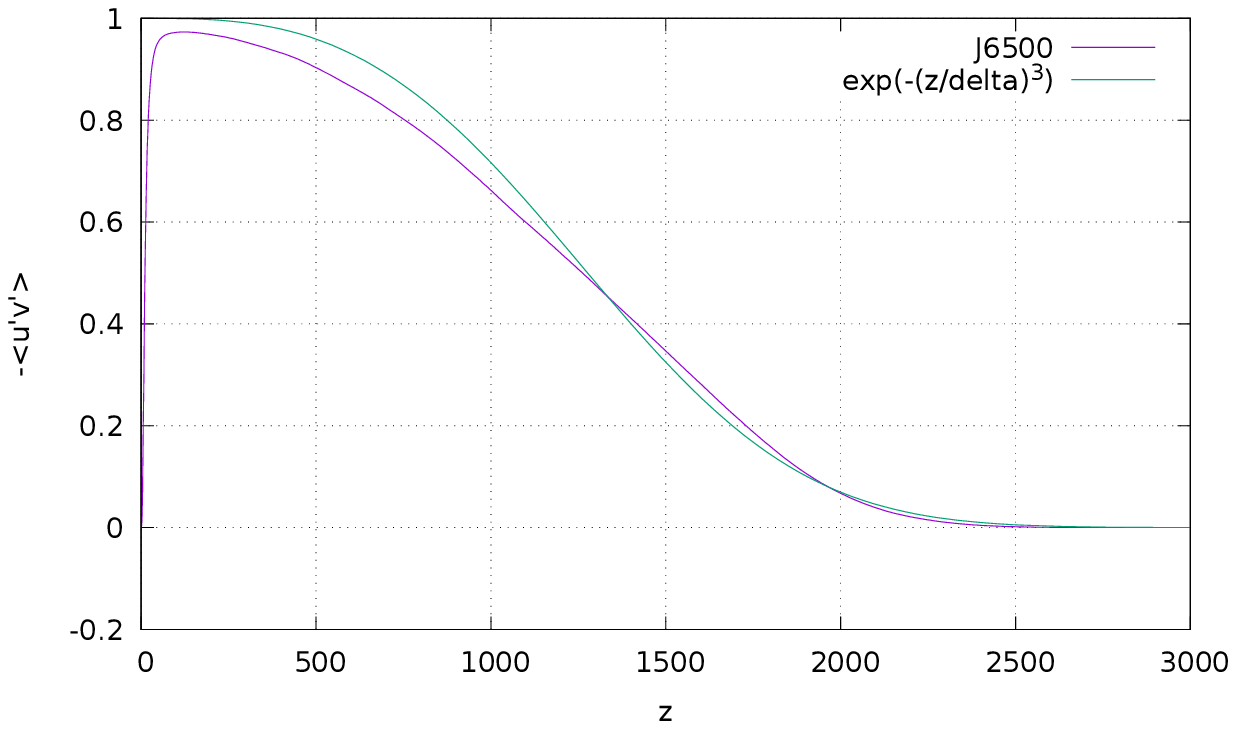}
\caption{Profile of the Reynolds stress $-\left< u'v'\right>$ in the DNS of \cite{Jimenez4} at $\Re_\theta=6500$, and its fit with a cubic exponential.}
\label{outeruv}
\end{figure}
In further support of this law of asymptotic decay, Figure \ref{outeruv} depicts the profile of Reynolds stress $-\left< u'v'\right>$ (whose outer value in wall units tends to $1$ as $z/\delta\rightarrow 0$) compared to the exponential $\e^{-(z/\delta)^3}$. That this graph too fits reasonably well the same exponential might partially be expected, on the basis of the ties of Reynolds stress with the velocity gradient (viscous stress) in the parallel momentum equation, but it is not totally obvious if it is remembered that in the boundary layer there is an additional convective, thickness-growth, term and that in the laminar regime this term provides the contribution that balances lateral diffusion; in the turbulent boundary layer, longitudinal thickness growth is even stronger. Therefore Figures \ref{notgaussian} and \ref{outeruv}, together, lend some credibility to the admittedly bold but fascinating conjecture that the exponential $\e^{-(z/\delta)^3}$, and not $\e^{-(z/\delta)^2}$ or some other exponent, provides the appropriate asymptotic decay of turbulent-boundary-layer mean quantities into the irrotational outer stream.
 
There is hardly a theory (or perhaps too many theories) to compare this conjecture to, but we can  recall that in the laminar boundary layer (and more generally in all diffusion-dominated phenomena), approach to the constant outer state is always like a gaussian $\e^{- (z/\delta)^2}$ (possibly multiplied by a small algebraic power depending on the precise quantity we are looking at); therefore the change in exponent denotes a stark departure from diffusive behaviour. Whether faster-than-diffusive mixing, a known generic property of turbulence, should a priori produce a faster-than-laminar or slower-than-laminar decay of the velocity defect is again not obvious; the present empirical observation of a faster decay is somewhat surprising to be found in a region characterized by entrainment and intermittency in its time evolution, but it should be remembered that the spatial decay of the average is not necessarily the same as the spatial decay of instantaneous quantities. Having a candidate exponent for such decay, nonetheless, might open the roadway to new and exciting interpretations in the future.

\end{document}